\documentclass[pra,reprint,amsmath,showpacs]{revtex4-1}

\usepackage[utf8]{inputenc}  
\usepackage[ngerman,american]{babel}
\usepackage{times}
\usepackage[T1]{fontenc}
\usepackage{amsmath, amssymb,amsfonts,mathtools}
\usepackage{graphicx}

\usepackage[normalem]{ulem}

\usepackage{braket}

\usepackage{hyperref}

\usepackage{color}

\begin{document}

\title{Robust population transfer in atomic beams induced \ by Doppler
shifts }
\author{R. G. Unanyan}
\affiliation{Technical University of Kaiserslautern, D-67663 Kaiserslautern, Germany}
\pacs{PACS number}

\begin{abstract}
The influence of photon-momentum recoil on adiabatic population transfer in
an atomic three-level lambda system is studied. It is shown that the Doppler
frequency shifts, due to atomic motion can play an important role in
adiabatic population transfer processes of atomic internal states by a pair
of laser fields. For the limiting case of slow atoms ( Doppler shift much
smaller than the photon recoil energy) the atoms occupy the same target
state regardless of the order of switching of laser fields, while for the
case of fast atoms interacting with the intuitive sequence of pulses, the
target state is the intermediate atomic state. It is shown that these
processes are robust with respect to parameter fluctuations, such as the
laser pulse area and the relative spatial offset (delay) of the laser beams.
The obtained results can be used for the control of temporal evolution of
atomic populations in atomic beams by externally adjustable Doppler shifts.
\end{abstract}

\date{\today }
\startpage{1}
\endpage{102}
\maketitle

\section{\protect\bigskip Introduction}

The transitions between atomic internal states induced by resonant laser
field is always accompanied by momentum transfer from photons to atoms. Many
elements of atom optics, e.g. deflectors and splitters, are based on this
effect \cite{Berman}. These elements separate the atomic wave function into
several components with different momentum \cite{Beams}- \cite{Lawall}. A
common approach for implementing such elements is to apply laser pulses with
suitable temporal pulse areas (e.g. so called $\pi $ or $\pi /2$ pulses).
This technique, however, is not robust because it requires a carefully
controlled duration and power of the pulse in order to assure\textbf{\ }the
desired area. An important paper by Marte, Zoller and Hall \cite{Zoller}
showed that an atomic-beam deflection by the stimulated Raman adiabatic
passage (STIRAP) technique \cite{Bergmann1990} can be implemented via
stimulated Raman transitions induced by counterpropagating \textbf{\ }laser
beams. Later on this idea was demonstrated independently by Lawall et al. 
\cite{Lawall} and Goldner et al. \cite{Goldner}. Those publications
stimulated the recognition of STIRAP, which was initially developed for
efficient excitation of molecules. Later, Theuer et al. \cite{Theuer}
developed a laser controlled variable beam-splitter based on the
tripod-STIRAP scheme \cite{Unanyan}, by varying the spatial overlap of
lasers.

For more details concerning current theoretical, experimental
developments and applications of STIRAP, the interested readers are referred
to the review articles \cite{STIRAP}. For completeness, however, we give a
brief description of STIRAP. The underlying physical mechanism of STIRAP is
the existence of an adiabatically decoupled, or dark state \cite{Arimondo},
which at early times coincides with the initial state \ and at late times is
aligned with the target state. When the pump and Stokes frequencies together
maintain two-photon resonance, then the only conditions to be fulfilled for
successful transfer are those of adiabaticity \cite{STIRAP},\cite{Shore},
i.e. large pulse area, and counterintuitive application of the pulses, i.e.
Stokes before pump.

The schemes suggested in Refs. \cite{Zoller} and \cite{Theuer} operate in
the regime where Doppler shifts of atomic transition frequencies due to the
photon recoil and the recoil energy are small compared to the Rabi\textbf{\ }%
frequencies and the transit-time broadening. In this limit the combined
system of atom plus fields is equivalent to a lambda or tripod atomic system
in which internal states can be labeled by different photon recoil momenta.
The idea of the beam splitters is then very clear: a population transfer
between these internal states via STIRAP \cite{Bergmann1990} or via
tripod-STIRAP \cite{Unanyan} necessarily means photon-momentum transfer to
the particles in the atomic beam. Thus in this regime, because of the
darkness of the state, the photon-recoil momentum does not play a role in
the evolution of internal atomic states\textbf{.}

It should be noted that standing wave laser beams (see e.g. \cite{Shore1981a}
and references therein) offer more efficient deflection of atoms than those
with traveling wave beams as in Refs. \cite{Zoller}- \cite{Theuer}. Our
goal, however, will not be to find an efficient way to deflect atoms,
rather, we ask a different question, namely whether is it possible to
control the dynamics of atomic internal states in a \textit{robust }\ way%
\textit{\ }via the velocity-induced Doppler shift of the atomic transition
frequency? To our knowledge, the problem with such formulation within the
context of robustness has never been discussed before. The results of the
present article show that due to Doppler  shifts the Landau-Zener transition 
\cite{Landau}   may occur between  the bright  and intermediate atomic
states. In the bare atomic basis, for the case when laser beams are ordered
intuitively i.e. the bright state is populated, this transition corresponds
to a robust population transfer from the initial to intermediate atomic
state. 

\section{Model}

Let us consider a monoenergetic atomic beam, crossing two laser beams but
not necessarily at a right angle. In particular, we consider an atomic
system with three levels coupled by two counterpropagating optical fields
with the same frequency $\omega $ and wave vectors $k$ whose parallel
propagation axes are spatially shifted. The geometry of the interaction
between atoms and lasers and  the atomic level scheme are drawn in Fig.~\ref%
{fig:Lambda}. 
\begin{figure}[h]
\includegraphics[width=0.8\linewidth]{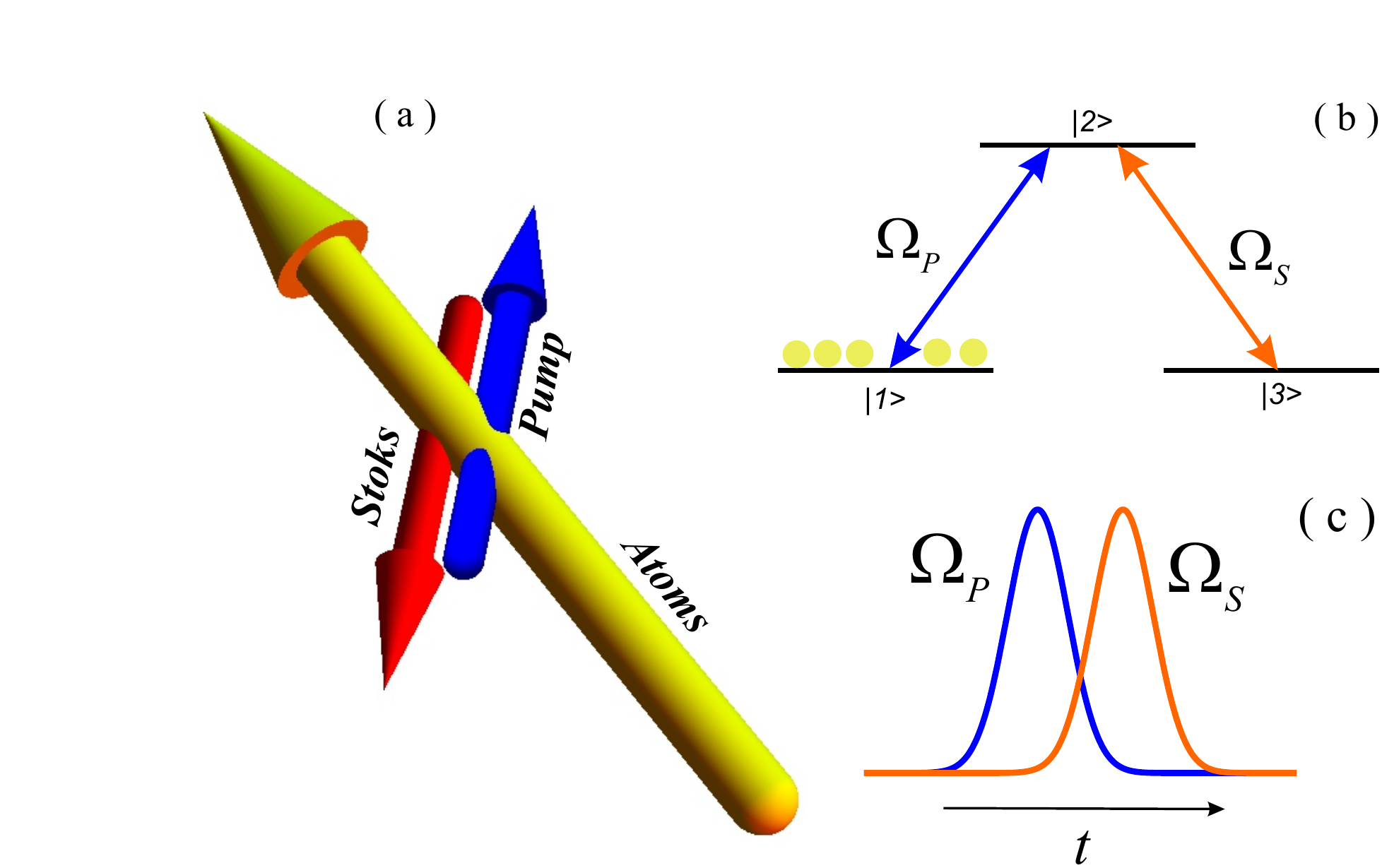}
\caption{(Color online) (a) Sketch of the geometry of the interaction
between atoms with two counterpropagating laser fields. (b) Three level
system with initial population in state $\left\vert 1\right\rangle $ (c)
Pulses in the intuitive sequence, the pump pulse precedes the Stokes pulse.}
\label{fig:Lambda}
\end{figure}

The sequence of interaction of atoms with laser beams is controlled by the
spatial displacement of their axes. The initial atom velocity in the
direction of the laser beams can be varied by changing the intersection
angle $\alpha $ between atomic velocity $\mathbf{v}$ and laser beams $%
\mathbf{k}$. The motion of atoms causes a Doppler shift $\sim \mathbf{v}%
\cdot \mathbf{k}$. The Doppler shift is large for small crossing angles $%
\alpha \approx 0$ and for large crossing angles $\alpha $ close to $\pi /2$
it tends to zero. As we will see later on, the possibility of controlling
the Doppler shift could provide a new way for controlling adiabatic
evolution processes of internal states of atoms.

The laser fields are detuned from the atomic transition by the single photon
detuning $\Delta =\omega _{21}-\omega =\omega _{23}-\omega $ . In what
follows, we will consider only the case of exact one-photon resonance, $%
\Delta =0$. The generalization to nonzero detuning is straightforward, and
requires one only to replace the recoil energy 
\begin{equation}
E_{r}=\frac{\hbar ^{2}k^{2}}{2M}  \label{recoil}
\end{equation}%
with $E_{r}+\Delta $. \ We assume that the atom-laser interaction time is
much shoter than the spontaneous lifetime of state $\left\vert
2\right\rangle $. \ This condition can be fulfilled, for example, for fast
atomic beams.

The atoms are described by atomic flip operators $\sigma _{nm}=\left\vert
n\right\rangle \left\langle m\right\vert ,$ $n,m=1,2,3$. The Hamiltonian for
the atom-field coupling along the $z$ axis ($z$ is directed along the
propagation direction of the one of lasers) in rotating wave approximation
has the following form \cite{Zoller} 
\begin{equation}
H=-\frac{\hbar ^{2}}{2M}\frac{\partial ^{2}}{\partial z^{2}}
\label{Schrodinger11}
\end{equation}%
\begin{equation*}
+\frac{\hbar \Omega _{P}\left( t\right) }{2}e^{-ikz}\sigma _{12}+\frac{\hbar
\Omega _{S}\left( t\right) }{2}e^{+ikz}\sigma _{32}+\text{H.c.},
\end{equation*}%
where $M$ is the mass of the atom. $\Omega _{P}\left( t\right) $ and $\Omega
_{S}\left( t\right) $ are space-independent, time-delayed Rabi frequencies
of the pump and Stocks fields, respectively.\textbf{\ }We have omitted the
transverse part of the kinetic energy since it is a conserved quantity.
Using the gauge transformation%
\begin{equation}
G=\exp \left[ ikz\left( \sigma _{11}-\sigma _{33}\right) \right] =
\label{Gauge_Transformation}
\end{equation}%
\begin{equation*}
\sigma _{22}+\sigma _{11}\exp \left( ikz\right) +\sigma _{33}\exp \left(
-ikz\right) 
\end{equation*}%
the Hamiltonian (\ref{Schrodinger11}) can be transformed to a more
convenient form%
\begin{equation}
H\rightarrow G\cdot H\cdot G^{-1}\rightarrow   \label{Schrodinger22}
\end{equation}%
\begin{equation*}
=\frac{\hbar k}{M}\cdot p\left( \sigma _{11}-\sigma _{33}\right) -\frac{%
\hbar ^{2}k^{2}}{2M}\sigma _{22}
\end{equation*}%
\begin{equation*}
+\frac{\hbar \Omega _{P}\left( t\right) }{2}\sigma _{12}+\frac{\hbar \Omega
_{S}\left( t\right) }{2}\sigma _{31}+\text{H.c}
\end{equation*}%
Here the important terms containing the constants of motion have been
omitted. Note that the atomic momentum commutes with the transformed
Hamiltonian. The momentum operator then is no longer a dynamical variable in
this frame\ and it has been replaced by a real parameter $p$. We see that
the states $\left\vert 1\right\rangle $ and $\left\vert 3\right\rangle $ are
shifted from the two photon resonance condition by the Doppler detunings $%
\frac{\hbar kp}{M}$. The energy of the state $\left\vert 2\right\rangle $ is
shifted by the recoil energy $\frac{\hbar ^{2}k^{2}}{2M}$.

The state vector $\left\vert \Psi \right\rangle $ transforms under the gauge
transformation (\ref{Gauge_Transformation}) into $G\left\vert \Psi
\right\rangle $. Throughout the paper, we assume that the initial state of
the system is $\left\vert \Psi \right\rangle =\exp \left( -\frac{i}{\hbar }%
p_{0}\cdot z\right) \left\vert 1\right\rangle $, where $p_{0}$ is the atom's
momentum projection onto the direction of the laser beams. We are interested
in the evolution of the system \ with the Hamiltonian (\ref{Schrodinger22})
when the Rabi frequencies $\Omega _{P}\left( t\right) $ and $\Omega
_{S}\left( t\right) $ are varied adiabatically in time.

\subsection{Adiabatic evolution}

To analyze adiabatic evolution of the system, it is convenient to introduce
dark, bright and atomic states by the following orthogonal transformation 
\begin{equation}
U\left( t\right) =\left( 
\begin{array}{ccc}
\cos \theta \left( t\right) & 0 & -\sin \theta \left( t\right) \\ 
0 & 1 & 0 \\ 
\sin \theta \left( t\right) & 0 & \cos \theta \left( t\right)%
\end{array}%
\right) .  \label{Transformation}
\end{equation}%
The mixing angle $\theta \left( t\right) $ is defined by%
\begin{equation}
\tan \theta \left( t\right) =\frac{\Omega _{P}\left( t\right) }{\Omega
_{S}\left( t\right) }.  \label{mixing_angle}
\end{equation}%
In the adiabatic limit of slowly changing mixing angle, 
\begin{align}
\frac{d\theta \left( t\right) }{dt}& <<\Omega _{\text{eff.}}\left( t\right) ,
\label{adiabatic_condition} \\
\Omega _{\text{eff.}}\left( t\right) & =\sqrt{\Omega _{P}^{2}\left( t\right)
+\Omega _{S}^{2}\left( t\right) },  \label{effective_Rabi}
\end{align}%
the corresponding Hamiltonian transforms into (for sake of simplicity we
omit below the argument $t$). 
\begin{equation}
H=H_{DD}+H_{D\overline{D}}+H_{\overline{D}D}+H_{\overline{D}\overline{D}},
\label{Hamiltonian_Dark_Bright}
\end{equation}%
where%
\begin{eqnarray}
H_{DD} &=&\frac{\hbar kp}{M}\left\vert D\right\rangle \left\langle
D\right\vert \cos 2\theta ,  \label{Dark} \\
H_{D\overline{D}} &=&H_{\overline{D}D}^{\dagger }=\frac{\hbar kp}{M}%
\left\vert D\right\rangle \left\langle B\right\vert \sin 2\theta ,
\label{Coupling} \\
H_{\overline{D}\overline{D}} &=&-E_{r}\sigma _{22}-\frac{\hbar kp}{M}%
\left\vert B\right\rangle \left\langle B\right\vert \cos 2\theta
\label{Bright}
\end{eqnarray}%
\begin{equation*}
+\frac{\hbar \Omega _{\text{eff.}}}{2}\left\vert B\right\rangle \left\langle
2\right\vert +\text{H.c}
\end{equation*}%
and the new orthogonal basis vectors are the dark $\left\vert D\right\rangle 
$ and bright $\left\vert B\right\rangle $ states defined in the following way%
\begin{equation}
\left\vert D\right\rangle =\left\vert 1\right\rangle \cos \theta -\left\vert
3\right\rangle \sin \theta ,  \label{Dark_State}
\end{equation}%
and 
\begin{equation}
\left\vert B\right\rangle =\left\vert 1\right\rangle \sin \theta +\left\vert
3\right\rangle \cos \theta .  \label{Bright_State}
\end{equation}%
The structure of the Hamiltonian (\ref{Hamiltonian_Dark_Bright}) tells us
that the terms $H_{\overline{D}D}$ and $H_{D\overline{D}}$ induce
transitions from the dark state to its complement subspace (bright and
atomic states) and vice versa. A sufficient condition to neglect them is to
ignore the term $\frac{\hbar kp}{M}\sin 2\theta $ compared to 
\begin{equation}
\delta =\min \left[ \left\vert \frac{\hbar kp}{M}\cos 2\theta -\mu
_{1}\right\vert ,\left\vert \frac{\hbar kp}{M}\cos 2\theta -\mu
_{2}\right\vert \right]  \label{spectrum_Gap}
\end{equation}%
i.e.%
\begin{equation}
\frac{\hbar kp}{M}\frac{\sin 2\theta }{\delta }<<1,
\label{Coupling_Condition}
\end{equation}%
where $\mu _{1,2}$ are the eigenvalues of $H_{\overline{D}\overline{D}}$.
One can verify that for relatively large Rabi frequencies%
\begin{equation}
\Omega _{\text{eff.}}>>\frac{\hbar kp}{M}\sin 2\theta ,  \label{Large_Area}
\end{equation}%
(along with the assumptions $kp>0$ and $\min \left[ \mu _{1},\mu _{2}\right]
<\frac{\hbar kp}{M}\cos 2\theta <\max \left[ \mu _{1},\mu _{2}\right] $)
condition (\ref{Coupling_Condition}) is fulfilled and the dark state
decouples from the dynamics. By combining this condition with the condition
of the adiabaticity (\ref{adiabatic_condition}), we arrive at the sufficient
condition 
\begin{equation}
\Omega _{\text{eff.}}>>\max \left[ 1/T,\frac{\hbar kp}{M}\sin 2\theta \right]
,  \label{Adiabatic_Doppler}
\end{equation}%
where $T$ is a time characterizing the pulse duration. This condition
ensures that the systems remains either in the initial dark state or its
complement subspace. In particularly, if a counterintuitive pulse sequence
is applied, i.e. if $\Omega _{S}$ is switched on and off before $\Omega _{P}$%
, i.e. the mixing angle starts at zero and then increases to $\pi /2$, thuse
the system is initially in the dark state $\left\vert D\right\rangle $ will
remain in this state, while according to the Eq. (\ref{Dark_State}) the
initial state $\left\vert 1\right\rangle $ will transform, up to an
irrelevant phase (proporional to the Doppler shift), into $\left\vert
3\right\rangle $. This dynamics has been described in reviews \cite{STIRAP}.
The inverse gauge transformation (\ref{Gauge_Transformation}) will yield
atomic beam deflection due to the momentum exchange $2\hbar k$ between the
atoms and laser photons. As was mentioned in the introduction, in this case,
the Doppler shift does not play a role in dynamics of the system.

\section{Population dynamics}

\subsection{Slow atomic beams}

In this subsection, we consider the case of the initial momentum of the
atoms $p_{0}$ in the direction of the laser beam being much smaller than the
recoil momentum $\hbar k$. This corresponds to the case when the atomic beam
is almost perpendicular to the laser beams. \ 

\begin{figure}[h]
\includegraphics[width=0.8\linewidth]{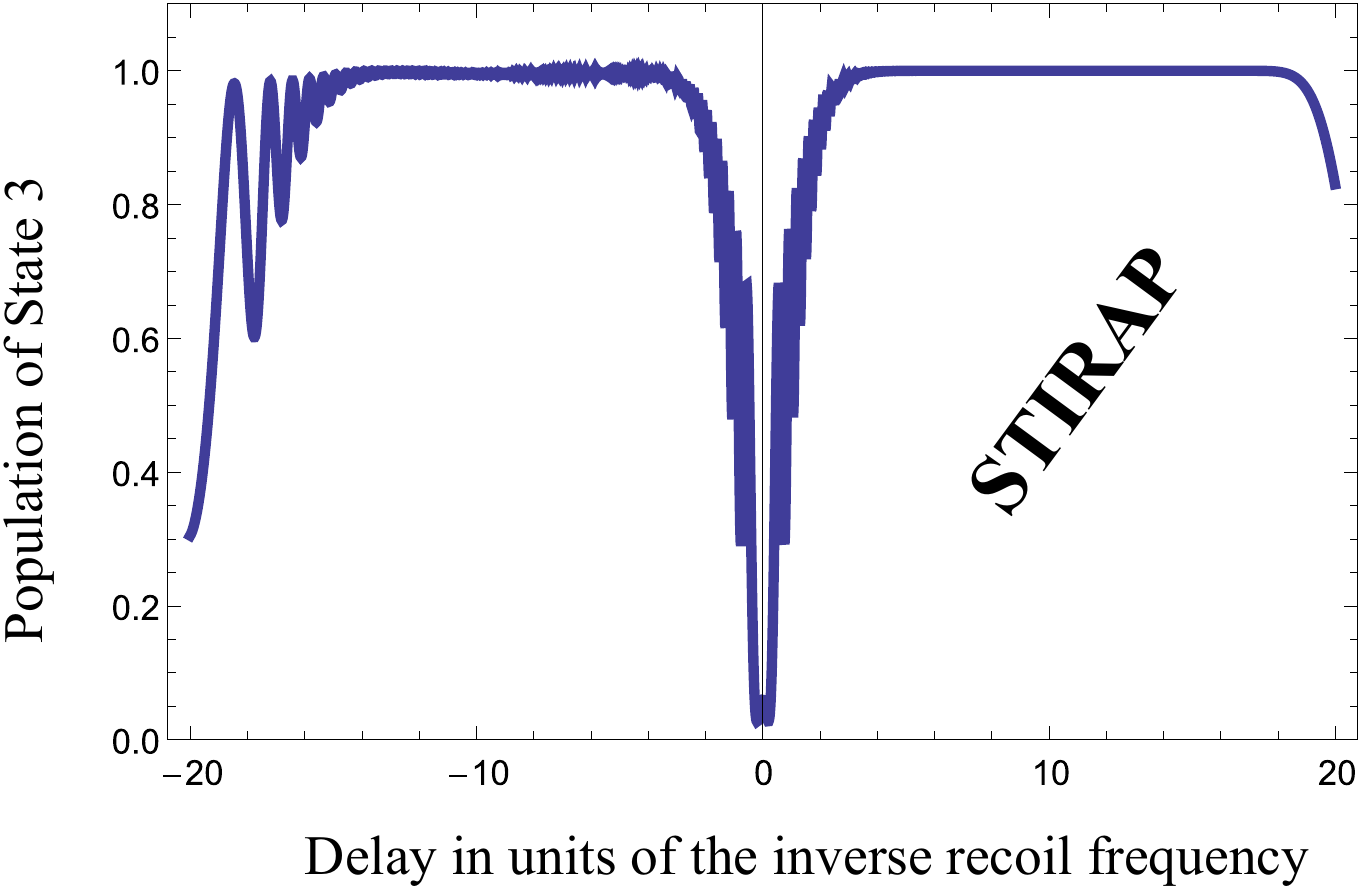}
\caption{(Color online) Population of the state $3$ as a function of the
time delay between pulses. The pulses are as in Eq. (\protect\ref{Pulses})
with equal pulse durations $T_{P}=T_{S}$ $=10$\ (time is measured in units
of the inverse recoil frequency), and equal areas $\Omega _{0}T=100$.
Positive values of the delay correspond to STIRAP. The initial state of the
atome is $\left\vert 1\right\rangle $ and $p_{0}=0.1\hbar k$.}
\label{fig:delay}
\end{figure}

\begin{figure}[h]
\includegraphics[width=0.8\linewidth]{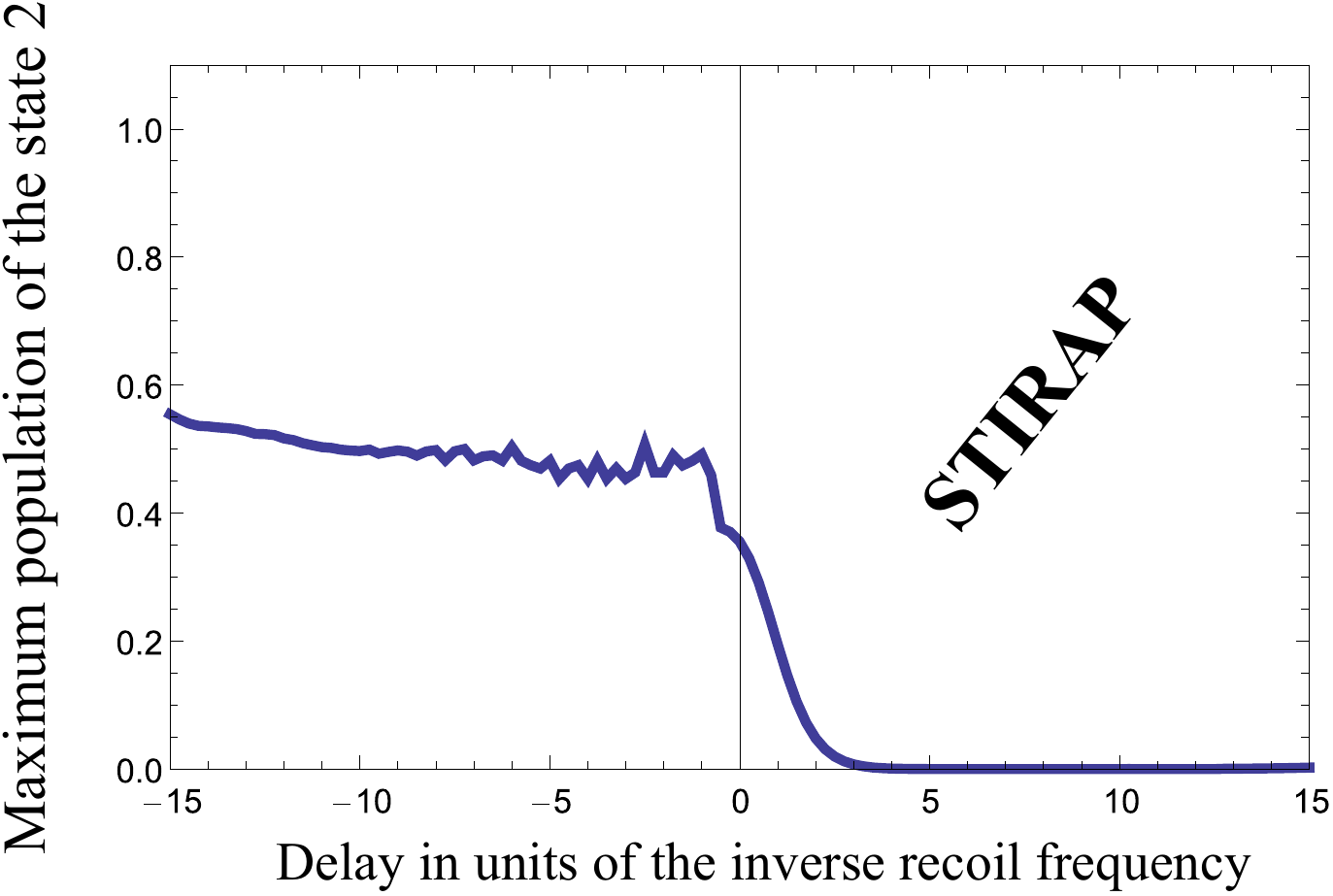}
\caption{(Color online) Maximum population of state $\left\vert
2\right\rangle $ during the evolution as a function of the time delay
between the pulses. Other parameters and units are the same as in Fig. ~%
\protect\ref{fig:delay}.}
\label{fig:maximum2}
\end{figure}

Fig.~\ref{fig:delay} shows the final atomic population of the state $%
\left\vert 3\right\rangle $ as a function of the delay between the coupling
pulses. This figure was obtained by numerically integrating the Schr\"{o}%
dinger equation with the full Hamiltonian (\ref{Hamiltonian_Dark_Bright}). 
The pulses have Gaussian shape 
\begin{eqnarray}
\Omega _{P}\left( t\right)  &=&\Omega _{0}\exp \left[ -\left( \frac{t-\tau }{%
T}\right) ^{2}\right] ,  \label{Pulses} \\
\Omega _{S}\left( t\right)  &=&\Omega _{0}\exp \left[ -\left( \frac{t+\tau }{%
T}\right) ^{2}\right]   \notag
\end{eqnarray}%
with equal pulse durations $T_{P}=T_{S}$ $=10$\ (in units of the inverse
recoil frequency), and equal areas $\Omega _{0}T=100$. The initial momentum
of the atoms is $p_{0}=0.1\hbar k$. The atoms are prepared initially in the
state $\left\vert 1\right\rangle $. Positive values of the delay correspond
to STIRAP, negative delays to an intuitive pulse sequence; i.e., the pulse $%
\Omega _{P}$ precedes $\Omega _{S}$. For both positive and negative delays,
we observe an efficient population transfer from the initial state $%
\left\vert 1\right\rangle $ to the state $\left\vert 3\text{ }\right\rangle $%
. The transfer efficiency approaches unity for a relative broad range of
delays. The case of STIRAP was discussed in \cite{Zoller}. In addition to
that, Fig.~\ref{fig:maximum2} shows that for negative delays in contrast to
ordinary STIRAP, the state $\left\vert 2\right\rangle $ is substantially
populated during the time evolution of the system. This indicates that the
transfer does not occur via adiabatic rotation of the dark state from $%
\left\vert 1\right\rangle $ to $\left\vert 3\right\rangle $.

The state vector for the initial problem can be found by the inverse gauge
transformation (\ref{Gauge_Transformation}) which leads to the change of the
atomic momentum from $p_{0}$ to $p_{0}+2\hbar k$. Hence, if the time scale
of the interaction is smaller than the radiative lifetime of state $%
\left\vert 2\right\rangle $, and the initial momentum of the atoms $p_{0}$
is much smaller than the recoil momentum, an intuitive pulse sequence also
transfers completely the population of the initial state $\left\vert
1\right\rangle $ to the state $\left\vert 3\right\rangle $. Hence, an atomic
beam deflection can be implemented by spatially shifted laser beams with
large pulse areas.

The mechanism of the population transfer for the negative delays (intuitive
sequence of pulses) can be understood qualitatively as follows: If an
intuitive pulse sequence is applied, the system starts its evolution from
the bright state $\left\vert B\right\rangle $, see: Eq. (\ref{Bright_State}%
). Due to the large recoil energy ($\hbar k>>p\cos 2\theta $), which plays
the role of a detuning, the bright state undergoes an adiabatic return
process under the condition 
\begin{equation}
E_{r}T>>\hbar .  \label{AdiabaticReturn}
\end{equation}%
Because of the chosen intuitive pulse order, the bright state transforms to
the bare atomic state $\left\vert 3\right\rangle $. This mechanism is
similar to the so called b-STIRAP process, \cite{Bergmann1990}, \cite%
{Vitanov1997},\cite{Halfmann}. The condition (\ref{AdiabaticReturn}),
however, might be questionable for atomic beams, because of the constraint
on the radiative lifetime. Indeed, it is important to recall that the
duration of the interaction of atoms with lasers should be shorter than the
upper state lifetime. Therefore, the condition $E_{r}\tau _{\text{sp}%
}>>\hbar $ should be satisfied, where $\tau _{\text{sp}}$ is the lifetime of
state $\left\vert 2\right\rangle $. In fact, the opposite situation occurs
in many experiments e.g. with noble atoms \cite{Vassen}.

\subsection{Fast atomic beams}

\begin{figure}[h]
\includegraphics[width=0.8\linewidth]{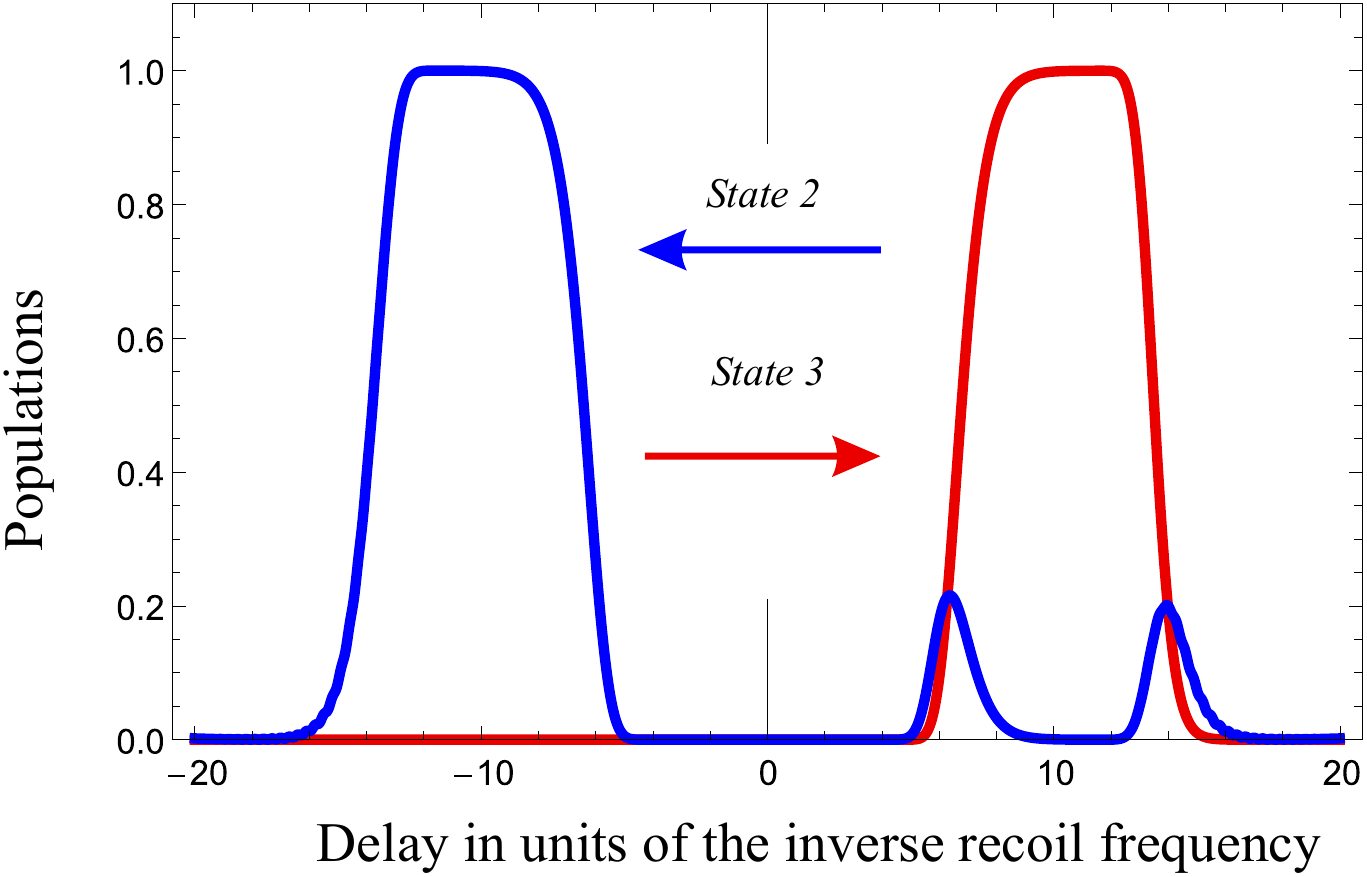}
\caption{(Color online) Final populations of states $\left\vert
2\right\rangle $ (red) and $\left\vert 3\right\rangle $ (blue) as a function
of the time delay between pulses. The initial momentum of atoms is $%
p_{0}=10\hbar k$. Other parameters and units are the same as in Fig. ~%
\protect\ref{fig:delay}.}
\label{fig:FastAtoms}
\end{figure}

\begin{figure}[h]
\includegraphics[width=0.8\linewidth]{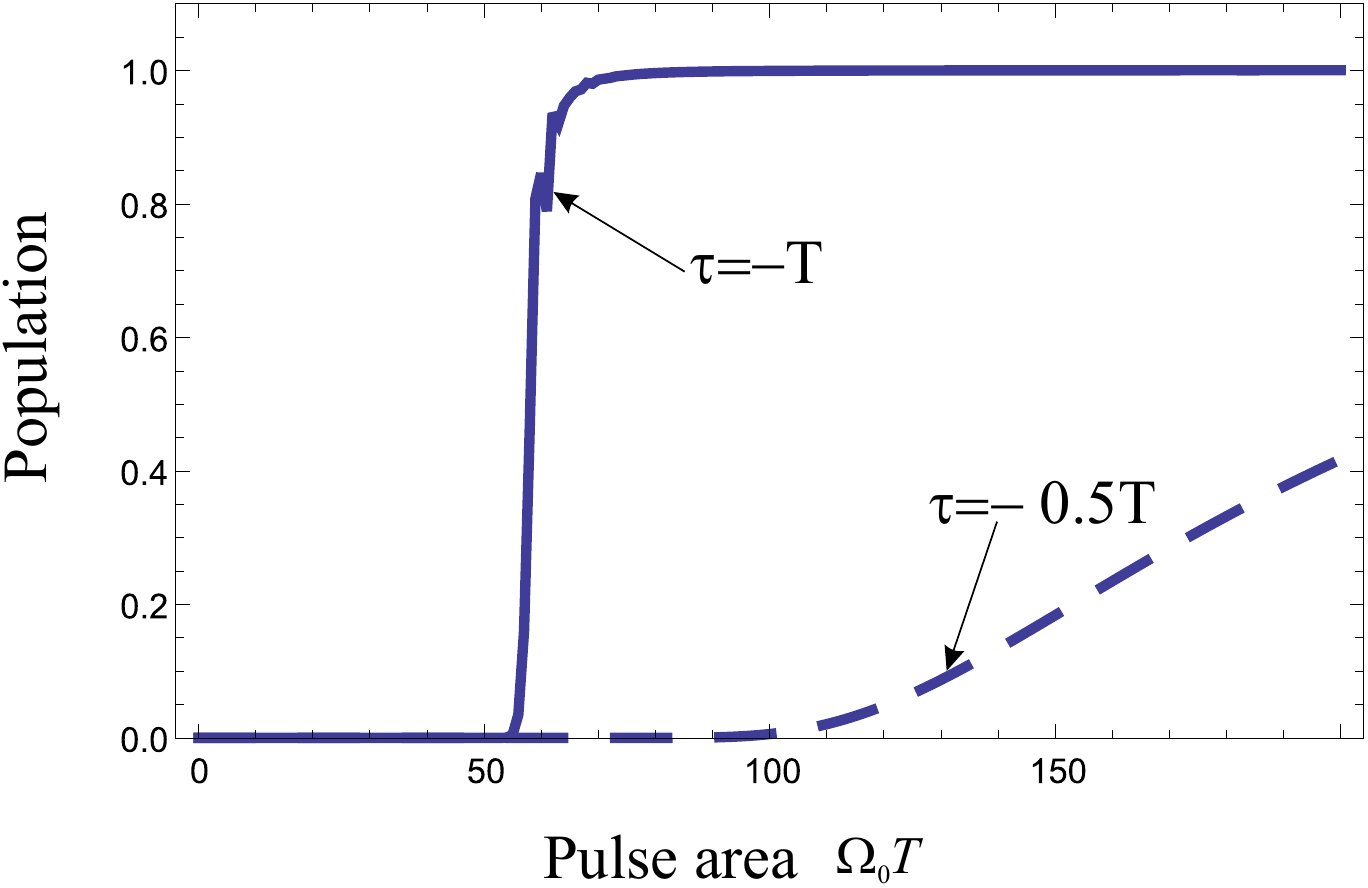}
\caption{(Color online) Final population of state $\left\vert 2\right\rangle 
$ under conditions of Fig. ~\protect\ref{fig:delay} as a \ function of $%
\Omega _{0}T$, for different delay time of pulses. The initial momentum $%
p_{0}=10\hbar k.$}
\label{fig:Area}
\end{figure}

Consider now the case where the initial momentum of the atom is larger than
the photon momentum. Fig.~\ref{fig:FastAtoms} shows the results of numerical
simulation for the final populations of states $\left\vert 2\right\rangle $
and $\left\vert 3\right\rangle $ for fast atoms ( $p_{0}=10\hbar k$)
depending on the time delay of pulses. Other parameters and units are the
same as in Fig.~\ref{fig:delay}. For positive large delays, the system ends
its evolution in state $\left\vert 3\right\rangle $ , i.e. the process is
STIRAP-like. The atom thus receives a momentum kick of $2\hslash k$. On the
other hand, for negative delays (around $0.8T$) the population transfer \
from $\left\vert 1\right\rangle $ to $\left\vert 2\right\rangle $ occurs
and, therefore, the momentum of the atom in the direction of laser beams
changes by $\hslash k$ . The transfer probability from $\left\vert
1\right\rangle $ to $\left\vert 2\right\rangle $ is robust for a wide range
negative delays. In Fig.~\ref{fig:Area} we show the dependence of the final
population in state $\left\vert 2\right\rangle $ for $\tau =T$ and $\tau =T/2
$ \ as function of the pulse area $\Omega _{0}T$. We thus see that the
adiabatic transfer to state $\left\vert 2\right\rangle $ is more efficient
for the case of largely delayed pulses. Fig. \ref{fig:Area} also shows that
there is a pronounced threshold area starting at $\Omega _{0}T\gtrsim 50$
for an efficient population transfer into state $\left\vert 2\right\rangle $%
. These observations are in full agreement with condition (\ref{Large_Area}).

In the following, we examine the dynamics of atomic populations in the case
of substantialy delayed laser pulses corresponding to small values of $\sin
2\theta $. In this case, condition (\ref{Large_Area}) is easy to satisfy for
small Rabi frequencies. Hence, the coupling between the dark and bright
components is negligible.

\ For a counterintuitive pulse sequence, the system is initially in a dark
state and the dynamics of the system is STIRAP-like. We, therefore, discuss
only the case when the Stokes and pump pulses are ordered intuitively, i.e.
the system starts \ its evolution from the bright state (\ref{Bright_State}%
). By neglecting the term (\ref{Coupling}) in Hamiltonian (\ref%
{Hamiltonian_Dark_Bright}), we arrive at the Landau--Zener Hamiltonian \cite%
{Landau} 
\begin{equation}
H=\left( E_{r}-\frac{\hbar k\cdot p}{M}\cos 2\theta \right) \left\vert
B\right\rangle \left\langle B\right\vert +\frac{\hbar \Omega _{\text{eff.}}}{%
2}\left( \left\vert B\right\rangle \left\langle 2\right\vert +\left\vert
2\right\rangle \left\langle B\right\vert \right) .  \label{TwoState}
\end{equation}%
It is easy to verify that the condition $p>\frac{\hbar k}{2}$ is essential
for the occurrence of a Landau-Zener like transition at the crossing point $%
\cos 2\theta =\frac{\hbar k}{2p}$. \ This condition with the adiabatic
condition $\Omega _{\text{eff.}}T>>1$ guarante a robust population transfer
from the bright to the atomic state. In this scheme, the atoms receive a
momentum kick only from the pump photons. The role of the Stokes laser is to
assist the Landau-Zener transition between the bright $\left\vert
B\right\rangle $ and $\left\vert 2\right\rangle $ states.

\section{ Conclusion and Discussion}

It is shown that the Doppler shift can play an important role in the
population transfer in three-level atoms driven by two counterpropagating
spatially shifted laser fields. In particularly, when the atoms interact
with intuitively ordered laser pulses, depending on the ratio of the initial
atomic and photonic momenta, the final atomic states are different. Namely,
for $p_{0}<<\hbar k$ \ and large recoil energies \ $E_{r}T>>1$ the atoms
occupy state $\left\vert 3\right\rangle $ and receive $2\hbar k$ momentum
regardless the order of switching the laser fields, while for the case of
fast atoms and an intuitive sequence of pulses, the target state is the
intermediate state $\left\vert 2\right\rangle $ and the corresponding
momentum kick is $\hbar k$. We showed that in the case of slow atoms and
intuitive sequence of laser pulses, the bright state, which is a linear
combination of the initial $\left\vert 1\right\rangle $ and target $%
\left\vert 3\right\rangle $ states, undergoes an adiabatic return process
(due to the large recoil energy). In the bare atomic basis, this process
corresponds to the $\left\vert 1\right\rangle $ $\rightarrow $ $\left\vert
3\right\rangle $ transition. Whereas in the second case (fast atoms), the
Landau-Zener transition occurs between the bright and atomic state $%
\left\vert 2\right\rangle $ and the system transforms from the initial state 
$\left\vert 1\right\rangle $ into state $\left\vert 2\right\rangle $.

For an experimental observation of these effects, the interaction time
between atoms and laser fields should be short enough $T<<\tau _{\text{sp.}}$%
. \ To fulfill this condition, a fast atom beam is required. For
implementing such a situation, a metastable cold helium beam with $^{3}S_{1}$
and $^{3}P_{1}$ transitions could be a possible candidate. Simple
calculations show that to avoid the spontaneous emission from the excited
state $^{3}P_{1}$ the helium atoms must enter the interaction region with
velocity in the range of $10$ to $100$ m$\cdot $s$^{-1}$. This is achievable
for metastable cold helium beams as was reported by Oberst et al. \cite%
{Oberst}. Another possible system which would be immune to the effects of
the spontaneous decay can be an ensemble of atoms with two lower stable
levels both coupled to a Rydberg state by laser fields \cite{Ott}. Atoms in
highly excited Rydberg states are remarkable stable against spontaneous
emission. The interaction between Rydberg atoms, however, could play an
important role for the evolution of atomic populations \cite{Molmer}. An
interesting extension of this work will be to study the influence of
atom-atom interactions on the effects considered above.

\bigskip

I am grateful to M. Fleischhauer, K. Bergmann, B.W Shore, N.V. Vitanov and
D. Petrosyan for many fruitful and stimulating discussions.

\end{document}